\documentclass[aps,pre,twocolumn,groupedaddress,showpacs,showkeys]{revtex4}

\usepackage{graphicx}
\usepackage{dcolumn}
\usepackage{bm}
\usepackage{amssymb}
\usepackage{amsmath}

\begin{document}

\title{TWO TYPES OF PHASE SYNCHRONIZATION DESTRUCTION}

\author{Alexander~E.~Hramov}
\author{Alexey~A.~Koronovskii}
\author{Maria~K.~Kurovskaya}
\affiliation{Faculty of Nonlinear
Processes, Saratov State University, Astrakhanskaya, 83, Saratov,
410012, Russia}

\date{\today}

\begin{abstract}
In this article we report that there are two different types of
destruction of the phase synchronization regime of chaotic
oscillators depending on the parameter mismatch as well as in the
case of the classical synchronization of periodic oscillators. When
the parameter mismatch of the interacting chaotic oscillators is
small enough the PS breaking takes place without the destruction of
the phase coherence of chaotic attractors of oscillators.
Alternatively, due to the large frequency detuning, the PS breaking
is accomplished by loss of the phase coherence of the chaotic
attractors.
\end{abstract}

\pacs{05.45.-a, 05.45.Xt, 05.45.Tp}

\keywords{coupled oscillators, chaotic synchronization, phase
synchronization, Lyapunov exponents}

\maketitle

\section*{Introduction}
\label{sct:Intro}

Phase synchronization (PS) of chaotic oscillators observed in many
systems (physical, technical, chemical, biological, physiological,
etc.) has attracted great attention of scientists
recently~\cite{Pikovsky:2002_SynhroBook,Anishchenko:2001_SynhroBook,Boccaletti:2002_ChaosSynchro}.
PS has practical
importance~\cite{Rosenblum:2001_HandbookBiologicalPhysics,Quiroga:2002_Kraskov},
with the phase synchronization concept involving the problems of the
phase coherence of chaotic
attractor~\cite{Rosenblum:2002_FrequencyMeasurement,Anishchenko:2004_ChaosSynchro}
and occurrence (destruction) of PS
regime~\cite{Rosenblum:1997_LagSynchro,Osipov:2003_3TypesTransitions}.

The scientists usually do not distinguish between types of the phase
synchronization regime destruction. As an exception we can mention
work~\cite{Osipov:2003_3TypesTransitions} where the existence of
three types of transitions to PS regime in the coupled oscillators
depending on the coherence properties of oscillations were
described. Contrary to the conclusion obtained
in~\cite{Osipov:2003_3TypesTransitions}, in this paper we report
that there are two different types of the PS regime destruction
depending on the detuning of the system control parameters as well
as in the case of the classical synchronization of the periodic
oscillators. We find the following: that when the parameter mismatch
of the interacting chaotic oscillators is small enough the PS
breaking takes place without destruction of the phase coherence of
chaotic attractors of oscillators, whereas for the large parameter
detuning with the decrease of the coupling strength between
oscillators (or the amplitude of the external signal) the PS
breaking is accompanied by the loss of the phase coherence of the
chaotic attractors. To make our arguments more convenient we compare
our methods and results with the ones described in
work~\cite{Osipov:2003_3TypesTransitions}.

The structure of this paper is the following.
Section~\ref{sct:Roesslers} presents two scenarios of the
destruction of the phase synchronization regime in the
unidirectionally coupled R\"ossler oscillators.
Section~\ref{sct:VdP} explains the mechanisms resulting in these
scenarios of the phase synchronization destruction.
Section~\ref{sct:LEs} shows the relationship between Lyapunov
exponents and the phase synchronization boundary.
Section~\ref{sct:RoesslersNonCoherent} displays the phase
synchronization regime destruction in the systems with initially
incoherent chaotic attractors. Section~\ref{sct:Conclusion} provides
the final conclusion.

\section{Destruction of phase synchronization regime in two unidirectionally coupled R\"ossler systems}
\label{sct:Roesslers}

Let us start our consideration with two unidirectionally coupled
R\"ossler systems
\begin{equation}
\begin{array}{l}
\dot x_{d}=-\omega_{d}y_{d}-z_{d},\\
\dot y_{d}=\omega_{d}x_{d}+ay_{d},\\
\dot z_{d}=p+z_{d}(x_{d}-c),\\
\\
\dot x_{r}=-\omega_{r}y_{r}-z_{r} +\varepsilon(x_{d}-x_{r}),\\
\dot y_{r}=\omega_{r}x_{r}+ay_{r},\\
\dot z_{r}=p+z_{r}(x_{r}-c),\\\end{array} \label{eq:Roesslers}
\end{equation}
where $\varepsilon$ is a coupling parameter. The control parameter
values have been selected by analogy
with~\cite{Aeh:2005_GS:ModifiedSystem,Harmov:2005_GSOnset_EPL} as
$a=0.15$, $p=0.2$, $c=10.0$. The $\omega_r$--parameter which
determines the main frequency of the response system has been
selected as $\omega_r=0.95$, and the analogous parameter $\omega_d$
of the drive system has been varied in the range from 0.8 to 1.1
providing the mismatch of the interacting oscillators. The mentioned
above control parameter values provide the phase coherence of
chaotic attractors of both drive and response systems for zero
coupling strength.

\begin{figure}[tb]
\centerline{\includegraphics*[scale=0.4]{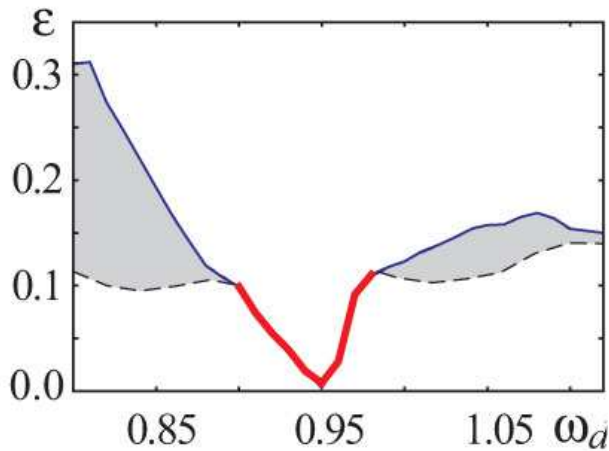}} \caption{(Color
online) The PS area on the $(\omega_d,\varepsilon)$-plane. The PS
boundary is shown in the range of the small parameter mismatch (the
bold solid line) of coupled oscillators as well as in the range of
the large one (the thin solid line). The chaotic attractor is
incoherent in the region shown in grey color. The lower bound of the
attractor non-coherence area is shown by the dashed line}
\label{fgr:RsslrParamPlane}
\end{figure}

In Fig.~\ref{fgr:RsslrParamPlane} the phase synchronization area on
the parameter plane $({\omega_d;\varepsilon})$ is shown. Following
to Ref.~\cite{Osipov:2003_3TypesTransitions} we use two criteria to
detect the presence of the phase synchronization regime. First, in
the regime of the phase synchronization, locking of the mean
frequencies
\begin{equation}
\Omega_d=\left\langle
\frac{d\varphi_{d}(t)}{dt}\right\rangle=\Omega_r=\left\langle
\frac{d\varphi_{r}(t)}{dt}\right\rangle
\label{eq:FrequencyLockingCondition}
\end{equation}
should take place, where $\langle\cdot\rangle$ denotes the time
average, $\varphi_{d}$ and $\varphi_{r}$ are instantaneous phases of
drive and response oscillators, respectively. Second, the presence
of PS may be detected by means of the examination of the phase
locking
condition~\cite{Pikovsky:2002_SynhroBook,Anishchenko:2001_SynhroBook,Boccaletti:2002_ChaosSynchro,%
Pikovsky:2000_SynchroReview}
\begin{equation}
|\varphi_{d}(t)-\varphi_{r}(t)|<\mathrm{const},
\label{eq:PhaseLockingCondition}
\end{equation}
with the instantaneous phase of chaotic signal being introduced in a
traditional way as the rotation angle $\varphi=\arctan(y/x)$.

Examining the behavior of two unidirectionally coupled R\"ossler
systems~(\ref{eq:Roesslers}) we have found two different types of
the PS regime destruction. When the parameter mismatch of the
interacting chaotic oscillators is large enough, the PS breaking is
accompanied by the loss of the phase coherence of the chaotic
attractors, whereas for the small parameter detuning with the
decrease of the coupling strength between oscillators PS breaking
takes place without destruction of the phase coherence of chaotic
attractors of oscillators. According
to~\cite{Pikovsky:2002_SynhroBook,Boccaletti:2002_ChaosSynchro} the
attractor is phase coherent when the phase trajectory rotates around
the origin of the projection plane without crossing it.
Alternatively, the attractor is supposed to be incoherent.

\begin{figure}[tb]
\centerline{\includegraphics*[scale=0.5]{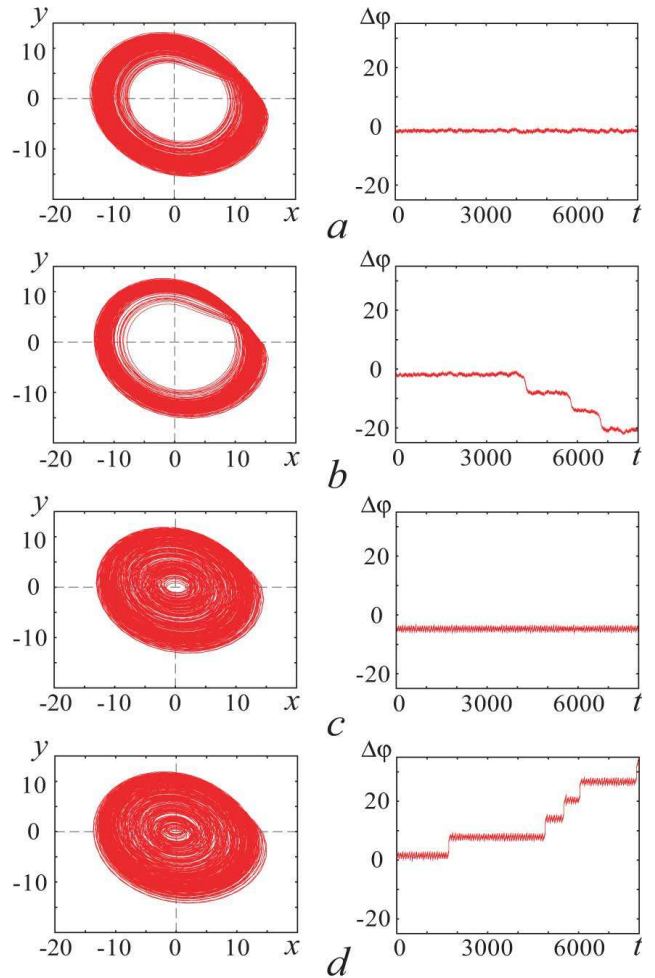}} \caption{(Color
online) Attractors of the response R\"ossler
system~(\ref{eq:Roesslers}) and the dependencies of the phase
differencies ${\Delta\varphi(t)=\varphi_d(t)-\varphi_r(t)}$ on time
$t$ for small detuning ($\omega_d=0.91$) of the drive R\"ossler
system (\textit{a}) above ($\varepsilon=0.08$) and (\textit{b})
below ($\varepsilon=0.075$) the onset of the PS regime. Analogous
figures for the large ($\omega_d=1.00$) detuning of the
$\omega_d$-parameter: (\textit{c}) $\varepsilon$ is above
($\varepsilon=0.127$) and (\textit{d}) below ($\varepsilon=0.123$)
the PS boundary} \label{fgr:RsslrRegimes}
\end{figure}

In Fig~\ref{fgr:RsslrRegimes},\,\textit{a,b} the chaotic attractors
of the response R\"ossler system and the dependence of the phase
difference ${\varphi_d(t)-\varphi_r(t)}$ on time $t$ are shown below
and above the onset of the PS regime for the small mismatch of
$\omega$-parameters. One can easily see that with the decrease of
the coupling strength $\varepsilon$ the PS regime is destroyed, with
the chaotic attractor remaining phase coherent. For the large
mistuning of the system control parameters the different scenario of
the PS destruction takes place (see
Fig.~\ref{fgr:RsslrRegimes},\,\textit{c,d}). Below the boundary of
the PS area the chaotic attractor of the response R\"ossler system
loses its phase coherence.

Since the notion of the attractor coherence plays the key role in
the chaotic synchronization theory the quantitative measure for the
phase coherence should be defined. The measure of coherence may be
characterized by means of the minimal distance $\rho$ between the
points of the phase trajectory and the origin
\begin{equation}\label{eq:CoherenceMeasure}
\rho=\min\limits_{t\rightarrow+\infty} \sqrt{x^2(t)+y^2(t)},
\end{equation}
where $x(t)$, $y(t)$ are the coordinates of the phase trajectory
projection. Obviously, $\rho >0$ for the phase coherent chaotic
attractor, and $\rho$ becomes equal to zero when attractor loses its
coherence.

In Ref.~\cite{Osipov:2003_3TypesTransitions} the coherence of
chaotic attractor of the system under study was characterized by the
sequence of the local maxima of $y(t)$, i.e. $\max(y)$. The chaotic
attractor is incoherent when some $\max(y)<Y_0$ occurs, where
$(X_0,Y_0)\approx(0,0)$ is the fixed point around which the phase
trajectory rotates. Such criterion is limited, since only one
coordinate is considered. Our definition~(\ref{eq:CoherenceMeasure})
of the coherence measure allows to generalize the approach used
in~\cite{Osipov:2003_3TypesTransitions} in such a way for both
coordinates to be taken into account.

\begin{figure}[tb]
\centerline{\includegraphics*[scale=0.4]{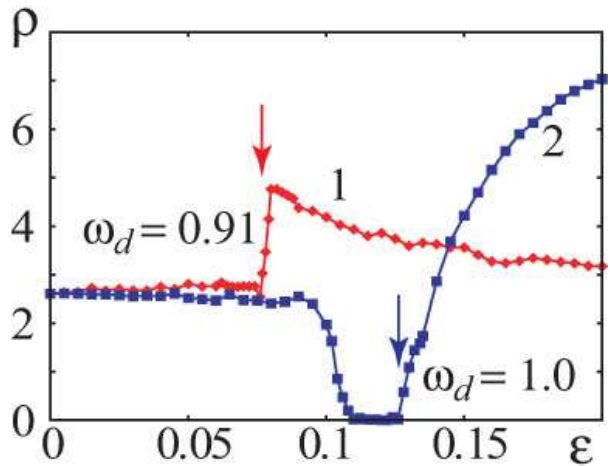}} \caption{(Color
online) The dependencies of the coherence measure $\rho$ on the
$\varepsilon$-parameter for the small (curve 1) and large (curve 2)
system parameter mismatches. The arrows show the values of the
coupling strength $\varepsilon$ corresponding to the PS
destruction.} \label{fgr:RsslrCoherenceMeasure}
\end{figure}

The dependencies of the coherence measure $\rho$ of the chaotic
attractor of the response system on the coupling strength
$\varepsilon$ are shown in Fig.~\ref{fgr:RsslrCoherenceMeasure} for
both small (curve~1) and large (curve~2) parameter mismatches.

One can easily see that for small detuning of the control parameters
(curve~1 in Fig.~\ref{fgr:RsslrCoherenceMeasure}) the measure of the
coherence $\rho$ is positive for all values of the coupling
strength, whereas for the large parameter mistuning (curve~2 in
Fig.~\ref{fgr:RsslrCoherenceMeasure}) there is an area just below
the PS boundary where the coherence measure is equal to zero.
Therefore, it is evident, that the chaotic attractor loses its
coherence when the PS regime is destroyed in the case of large
parameter detuning and remains phase coherent both below and above
the PS boundary if parameter mistuning is small enough.

The area on the plane $(\omega_d,\varepsilon)$ where the chaotic
attractor of the response system is incoherent is shown in
Fig.~\ref{fgr:RsslrParamPlane} in grey color. This region is joined
closely to the boundary of the phase synchronization regime for the
$\omega_d$-values detuned essentially from the $\omega_r$-parameter
of the response R\"ossler system. Therefore, we can come to
conclusion that depending on mistuning of the control parameters of
system~(\ref{eq:Roesslers}) the PS destruction occurs in two
different ways, the mechanisms resulting to these types of PS
destruction being supposed to be dissimilar.

Therefore, the different manifestations of these mechanisms may be
observed in the vicinity of the PS boundary. In particular, two
different types of the pre-transitional behavior revealed below the
PS boundary may be considered as an indication of different
mechanisms governing the scenarios of the phase synchronization
destruction. Indeed, it was found that the type-I intermittency and
the super-long laminar behavior (so-called ``eyelet
intermittency''~\cite{Pikovsky:1997_EyeletIntermitt}) take place for
small differences in the natural frequencies of the drive and
response systems~\cite{Pikovsky:1997_PhaseSynchro_UPOs,Lee:1998:PhaseJumps,%
Boccaletti:2002_LaserPSTransition_PRL,Rosa:1998_TransToPS,%
Pikovsky:1997_EyeletIntermitt}, while, as far as large values of the
natural frequency differences are concerned, the ring intermittency
emerges~\cite{Hramov:RingIntermittency_PRL_2006}.

Let us show that the same both types of the synchronization
destruction are also observed in the case of the driven periodic
oscillators. This topic is discussed in detail in the next section.


\section{Phase synchronization destruction of van der Pol oscillator}
\label{sct:VdP}

One of the approaches allowing to reveal different aspects of the
synchronization phenomenon is the consideration of the periodic
oscillators (see.,
e.g.~\cite{Balanov:2002_CRvsPS,Woafo:2002_SynchroDuration,Balanov:2005_UnstableTori,%
Hramov:2006_Prosachivanie}). Therefore, let us use the classical
model of the synchronization theory, namely, van der Pol oscillator
\begin{equation}
{\ddot{x}-(\lambda-x^2)\dot{x}+x=A\sin(\omega_et)}
\label{eq:VdPOscillator}
\end{equation}
driven by the external harmonic signal with the amplitude $A$ and
frequency $\omega_e$ to explain the mechanisms causing the PS regime
destruction in the coupled R\"ossler systems~(\ref{eq:Roesslers}).
The value of the control parameter has been selected as
$\lambda=0.1$. Oscillations in this case are certainly not chaotic
but we can use the concepts of phase synchronization and phase
coherence of an attractor in the same way as it is done for the
chaotic oscillators. The phase of van der Pol oscillator is
introduced as the rotation angle $\varphi=\arctan(y/x)$ as well as
for the R\"ossler systems~(\ref{eq:Roesslers}) considered above,
whereas the phase of the external signal has been calculated as
$\varphi_e(t)=\omega_et$.

\begin{figure}[tb]
\centerline{\includegraphics*[scale=0.4]{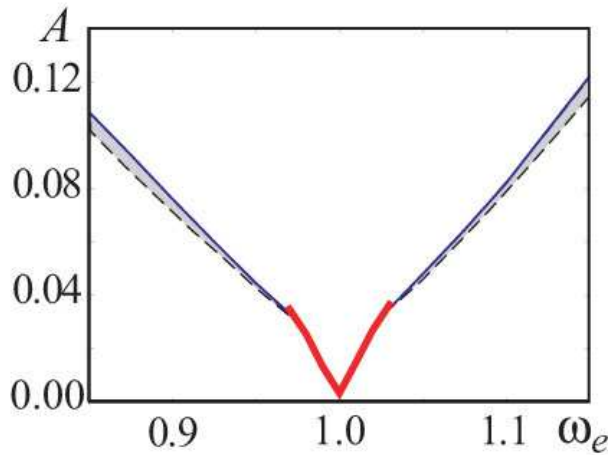}} \caption{(Color
online) The PS area on the $(\omega_e,A)$-plane for the driven van
der Pol oscillator. The PS boundary is shown in the range of the
small frequency mismatch (the bold solid line) as well as in the
range of the large one (the thin solid line). The attractor is
incoherent in the narrow region shown in grey color. The lower bound
of the attractor non-coherence area is shown by the dashed line}
\label{fgr:VdPParamPlane}
\end{figure}

\begin{figure}[tb]
\centerline{\includegraphics*[scale=0.5]{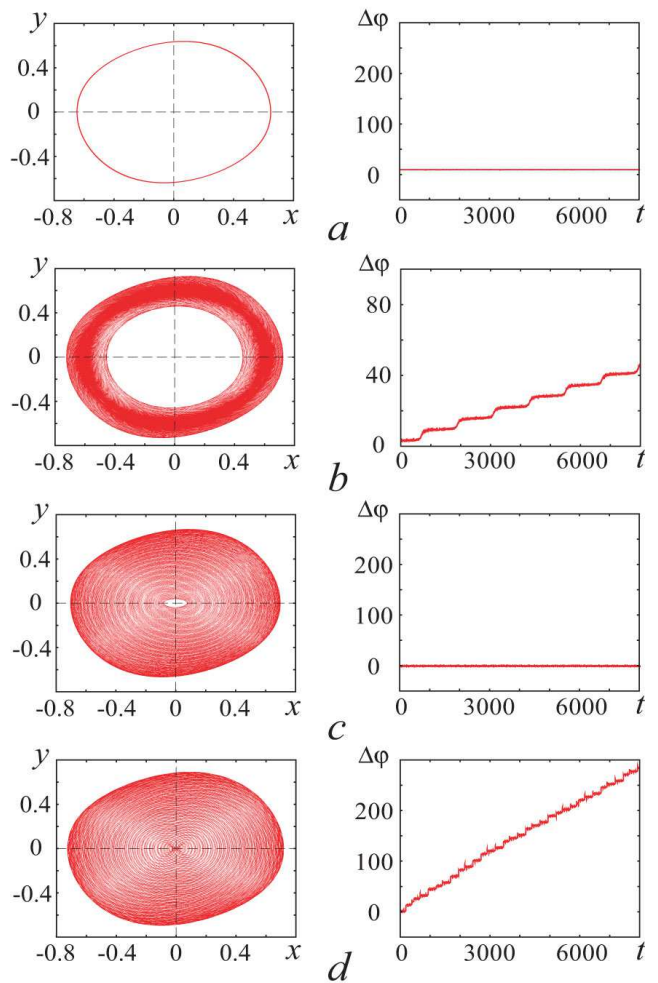}} \caption{(Color
online) Attractors of van der Pol oscillator driven by the external
harmonic signal~(\ref{eq:VdPOscillator}) and the dependencies of the
phase difference $\Delta\varphi(t)=\varphi(t)-\omega_et$ on time $t$
for small ($\omega_e=0.98$) frequency detuning (\textit{a}) above
($A=0.0250$) and (\textit{b}) below ($A=0.0230$) the onset of the PS
regime. Analogous figures for the large (${\omega_e=0.9}$) detuning
of the external signal frequency: (\textit{c}) $A$ is above
(${A=0.0775}$) and (\textit{d}) below ({A=0.0750}) the PS boundary}
\label{fgr:VdPRegimes}
\end{figure}

The synchronization area on the plane $(\omega_e;A)$ is shown for
the driven van der Pol oscillator~(\ref{eq:VdPOscillator}) in
Fig.~\ref{fgr:VdPParamPlane}. As well as for the unidirectionally
coupled R\"ossler systems~(\ref{eq:Roesslers}) described in
Sec.~\ref{sct:Roesslers}, two scenarios of the phase synchronization
destruction may be observed. These two scenarios are illustrated in
Fig~\ref{fgr:VdPRegimes}. The attractors of van der Pol oscillator
driven by the external harmonic signal~(\ref{eq:VdPOscillator}), and
the dependencies of the phase difference
${\Delta\varphi(t)=\varphi(t)-\varphi_e(t)}$ on time $t$ are shown
in Fig.~\ref{fgr:VdPRegimes},\,\textit{a,b} for the small
frequencies detuning ${(\omega_0-\omega_e)}$ below and above the
onset of the PS regime (here
${\omega_0=\langle\dot\varphi(t)\rangle\simeq 1.0}$ is the natural
frequency of the autonomous van der Pol oscillator). Evidently, with
the decrease of the amplitude of the external signal $A$ the PS
regime is destroyed, with the attractor remaining phase coherent.
Alternatively, for large mistuning of the frequencies, the different
scenario of the PS destruction takes place (see
Fig.~\ref{fgr:VdPRegimes},\,\textit{c,d}), since the attractor loses
its phase coherence below the boundary of the PS area.

\begin{figure}[tb]
\centerline{\includegraphics*[scale=0.4]{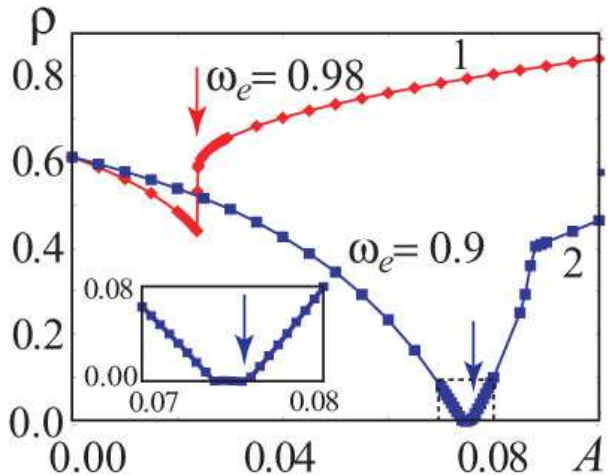}} \caption{(Color
online) The dependence of the coherence measure $\rho$ on the
$A$-parameter for the small (curve 1, $\omega_e=0.98$) and large
(curve 2, $\omega_e=0.9$) frequency mismatches. The arrows show the
values of $A$-parameter corresponding to the PS destruction. In the
frame the small part of the $\rho(A)$ dependence near the point of
the PS destruction for the large frequency detuning is shown}
\label{fgr:VdPCoherenceMeasure}
\end{figure}

This statement may also be confirmed by means of the dependencies of
the coherence measure $\rho$ of the attractor of the driven van der
Pol oscillator on the amplitude $A$ of the external signal given in
Fig.~\ref{fgr:VdPCoherenceMeasure} for both small and large
frequencies mismatches ${(\omega_0-\omega_e)}$. Again, as in the
case of coupled R\"ossler systems
(Fig.~\ref{fgr:RsslrCoherenceMeasure}) the measure of the coherence
$\rho$ is positive for all values of the amplitude of the external
signal (curve~1 in Fig.~\ref{fgr:VdPCoherenceMeasure}), if the
frequencies $\omega_0$ and $\omega_e$ are detuned slightly, i.e.,
the attractor of the driven van der Pol oscillator remains phase
coherent both below and above the PS boundary. As far as the large
frequency mistuning is considered (curve~2 in
Fig.~\ref{fgr:VdPCoherenceMeasure}) there is a narrow range located
just below the PS boundary on which the coherence measure $\rho$ is
equal to zero, hence the attractor loses its phase coherence, when
the PS regime is destroyed.

The area on the plane $(\omega_e,A)$, where the attractor of the
driven van der Pol oscillator~({\ref{eq:VdPOscillator}) is
incoherent, is shown in Fig.~\ref{fgr:VdPParamPlane} in grey color.
As in the case of R\"ossler systems~(\ref{eq:Roesslers}) considered
in Sec.~\ref{sct:Roesslers} this region is joined closely to the
boundary of the phase synchronization regime (although the width of
this area is less than in the case of R\"ossler oscillators) when
the frequency of the external signal is detuned essentially from the
natural frequency of van der Pol oscillator.

Thus, when the PS regime is destroyed, the evolution of the
attractor of van der Pol oscillator~(\ref{eq:VdPOscillator}) driven
by the external harmonic signal is the same as the evolution of the
attractor of the response R\"ossler~(\ref{eq:Roesslers}) system with
a decrease of the coupling strength $\varepsilon$, despite of the
fact that the R\"ossler system dynamics is chaotic, and van der Pol
oscillator one
--- periodic. Therefore, we can assume that the destruction of the
phase synchronization regime is governed by the same mechanisms in
both cases, with transition from synchronous to asynchronous state
of van der Pol oscillator being possible to be explained
analytically by means of the complex amplitude method.

It is well known that there are two different scenarios for
synchronization destruction in a periodic oscillator driven by an
external force,  corresponding to small and large detunings between
the natural and external signal frequencies (see, e.g.,
tutorial~\cite{Pikovsky:2000_SynchroReview}), respectively. Under
certain conditions (i.e., for the periodically forced weakly
nonlinear isochronous oscillator), the complex amplitude method may
be used to find the solution describing the oscillator behavior in
the form
\begin{equation}
{x(t)=\mathrm{Re}\,a(t)e^{i\omega t}}.
\label{eq:ComplexAmplitudeMethod}
\end{equation}
For the complex amplitude $a(t)$ one obtains the averaged
(truncated) equations
\begin{equation}
{\dot{a}=-i\nu a+a-|a|^2a-ik},
\end{equation}
where ${\nu}$ is the frequency mismatch, and $k$ is the
(renormalized) amplitude of the external force. For the small $\nu$
and large $k$ the stable fixed point on the plane of the complex
amplitude
\begin{equation}
{a^*=Ae^{j\phi}=\mathrm{const}}
\end{equation}
corresponds to the synchronous regime, with the synchronization
destruction corresponding to the local saddle-node bifurcation
associated with the global bifurcation of the limit cycle birth.
Note, in this case the amplitude of the originated limit cycle is
large initially. For the large frequency mismatches, the different
mechanism of synchronization destruction is observed. With the
decrease of $k$-value the fixed point (stable node) on the complex
amplitude plane becomes sequentially a stable focus and an unstable
focus (via the Andronov--Hopf bifurcation), with the limit cycle
originating with the infinitesimal amplitude. With the further
decrease of $k$-parameter, the amplitude of the limit cycle grows
from zero monotonically. In this case, the phase synchronization
destruction is connected with the limit cycle location on the
complex amplitude plane~\cite{Pikovsky:2000_SynchroReview}.  When
the limit cycle starts enveloping the origin, the synchronization
regime begins to be destroyed. Obviously, in this case the attractor
loses its phase coherence due to the fact that there are the moments
of time when the modulus of amplitude $|a(t)|$ is equal to zero. As
far as the small frequency mismatch is concerned, the attractor is
always phase coherent due to $|a(t)|\neq 0$, $\forall
t\in(-\infty;+\infty)$.

Since the use of Eq.~(\ref{eq:ComplexAmplitudeMethod}) for
application of the complex amplitude method is equivalent to the
transition to the revolving coordinate system, two mechanisms of the
phase synchronization destruction described above may be easily
revealed by means of the consideration of the behavior of the driven
oscillator on the plane $(x',y')$ rotating with the frequency
$\omega_e$ of the external signal around the origin. These
considerations of the rotating plane may be made apparent by using
the coordinate transformation
\begin{equation}
\begin{array}{l}
x'=x_r\cos\varphi+y_r\sin\varphi,\\
y'=-x_r\sin\varphi+y_r\cos\varphi,\\
\end{array}
\label{eq:CoordinateTransformation}
\end{equation}
where $\varphi=\varphi_e(t)$ is the instantaneous phase of the
external signal.

\begin{figure}[tb]
\centerline{\includegraphics*[scale=0.4]{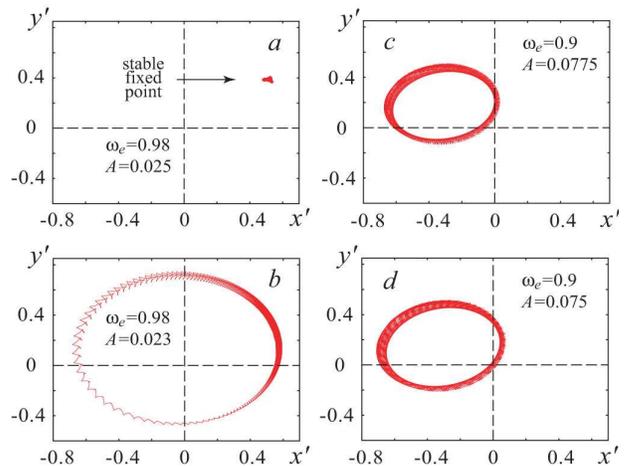}} \caption{(Color
online) The phase trajectories of driven van der Pol oscillator on
the rotating plane $(x',y')$ for the small (\textit{a,b}) and large
(\textit{c,d}) parameter mistuning, the $A$-values having been
selected just above (\textit{a,c}) and below (\textit{b,d}) the PS
boundary. Figures (\textit{a,c}) correspond to the synchronization
regime, figures (\textit{b,d}) illustrate the asynchronous behavior}
\label{fgr:VdPSynchroRotatingPlane}
\end{figure}

In case of small frequency detuning ${(\omega_0-\omega_e)}$ the
stable node is observed on the rotating plane $(x';y')$ for the
synchronized van der Pol oscillator
(Fig.~\ref{fgr:VdPSynchroRotatingPlane},\,\textit{a}). As soon as
the amplitude of the external signal starts to be below the phase
synchronization boundary, the local saddle-node bifurcation,
associated with the global bifurcation of the limit cycle birth,
takes place and on the $(x',y')$-plane the cycle with the initially
large amplitude is becoming observable
(Fig.~\ref{fgr:VdPSynchroRotatingPlane},\,\textit{b}) instead of the
stable node. As far as the large frequency difference
${(\omega_0-\omega_e)}$ is concerned a cycle (may be smeared, if the
$\lambda$-parameter is not small enough) should be found on the
rotating plane both below and above the PS synchronization boundary
(Fig.~\ref{fgr:VdPSynchroRotatingPlane},\,\textit{c,d}). With the
decrease of the intensity of the external signal the amplitude of
the cycle increases monotonically, with the destruction of the phase
synchronization regime corresponding to the control parameter value
when the cycle starts to cross the origin
(Fig.~\ref{fgr:VdPSynchroRotatingPlane},\,\textit{d}). When the
cycle does not envelop the origin (see
Fig.~\ref{fgr:VdPSynchroRotatingPlane},\,\textit{c}), the phase
synchronization regime takes place.

Obviously, if the mechanisms causing the PS regime destruction are
the same in the cases of both periodic and chaotic oscillators, one
has to obtain the similar results for chaotic oscillators, too.
Therefore, let us consider the dynamics of coupled R\"ossler
systems~(\ref{eq:Roesslers}) on the rotating
plane~(\ref{eq:CoordinateTransformation}) in the same way as it was
done for the driven van der Pol oscillator, with the phase
$\varphi=\varphi_d(t)$ of the drive R\"ossler system being used
instead of the phase of the harmonic signal.

\begin{figure}[tb]
\centerline{\includegraphics*[scale=0.4]{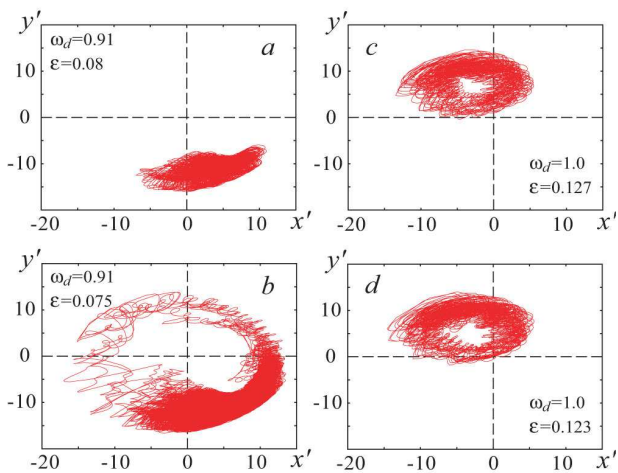}} \caption{(Color
online) The phase trajectories of the response R\"ossler
system~(\ref{eq:Roesslers}) on the rotating plane $(x',y')$ for
small (\textit{a,b}, $\omega_d=0.91$) and large (\textit{c,d},
$\omega_d=1.0$) parameter mistuning, the coupling strength values
having been selected just above (\textit{a,c}) and below
(\textit{b,d}) the PS boundary}
\label{fgr:RsslrSynchroRotatingPlane}
\end{figure}

The results obtained for R\"ossler systems~(\ref{eq:Roesslers}) are
presented in Fig.~\ref{fgr:RsslrSynchroRotatingPlane}. For the small
system parameter mismatch
(Fig.~\ref{fgr:RsslrSynchroRotatingPlane},\,\textit{a}) the behavior
of the synchronized response oscillator in the vicinity of the PS
boundary looks like noise smeared stable fixed point on the
$(x',y')$-plane. This effect arises insofar as the R\"ossler system
may be considered as a noise smeared periodic oscillator (see,
e.g.,~\cite{Pikovsky:1997PhaseSynchro}). When the coupling strength
between R\"ossler oscillators decreases and the PS regime is
destroyed, the stable noise smeared fixed point disappears and, in
full agreement with the dynamics of the driven periodic oscillator
(compare with
Fig.~\ref{fgr:VdPSynchroRotatingPlane},\,\textit{a,b}), the noise
smeared limit cycle with initially large amplitude occurs
(Fig.~\ref{fgr:RsslrSynchroRotatingPlane},\,\textit{b}). Similarly,
as well as in the case of van der Pol oscillator, if the large
parameter mismatches are concerned, the noise smeared cycle is
observed both below and above the PS regime boundary (see
Fig.~\ref{fgr:RsslrSynchroRotatingPlane},\,\textit{c,d} and compare
it with Fig.~\ref{fgr:VdPSynchroRotatingPlane},\,\textit{c,d}), with
the destruction of the phase synchronization regime corresponding to
the cycle starting enveloping the origin
(Fig.~\ref{fgr:RsslrSynchroRotatingPlane},\,\textit{d}).

Thus, taken into account the results mentioned above we can draw a
conclusion that the mechanisms of two types of the chaotic PS
destruction are the same as in the case of the synchronization of
periodic oscillations.

\section{Lyapunov exponents and phase synchronization boundary}
\label{sct:LEs}

Let us discuss now the problem concerning the relationship between
the observed types of PS destruction and transitions to PS regime
described in Ref.~\cite{Osipov:2003_3TypesTransitions}.
In~\cite{Osipov:2003_3TypesTransitions} three different types of
transitions from asynchronous dynamics to the phase synchronization
regime were reported of,  depending on the coherence properties of
the chaotic attractors of the uncoupled systems.

The degree of noncoherence of attractors of the autonomous
oscillators was evaluated in~\cite{Osipov:2003_3TypesTransitions} by
means of the phase diffusion coefficient
\begin{equation}
\langle(\varphi(t)-\langle\varphi(t)\rangle)^2\rangle=2D_\varphi t,
\end{equation}
where $\varphi(t)$ is the instantaneous phase of chaotic oscillator,
$\langle\cdot\rangle$ denoting the ensemble average. For a phase
coherent chaotic attractor the phase increases approximately
uniformly and the value of $D_\varphi$ is rather small, whereas for
a funnel attractor the increase of the phase is strongly nonuniform
and $D_\varphi$ is a few orders larger in magnitude.

As a criterion of distinction of different types of transitions
to/from the PS regime, the correlation between the dependency of the
PS synchronization threshold (the critical curve $l_1$) and the
dependencies of Lyapunov exponents (LEs) (the critical curve $l_2$
corresponds to the transition of one of the zero Lyapunov exponents
to the negative values and the critical curve $l_3$ corresponds to
passing through zero of the one of the positive Lyapunov exponents)
on the coupling strength has been used. The types of transitions
described in~\cite{Osipov:2003_3TypesTransitions} were the
following.

(I) Both uncoupled oscillators are characterized by the phase
coherent chaotic attractors, therefore the value of the diffusion of
phase $D_\varphi$ is small (e.g., $D_\varphi\sim10^{-3}$ for the
system considered in~\cite{Osipov:2003_3TypesTransitions}). In this
case, the phase synchronization regime is assumed to occur
immediately after the transition of one of the zero Lyapunov
exponents to a negative value, i.e., the critical curves $l_1$ and
$l_2$ are close to each other, curve $l_2$ preceding curve $l_1$.

(II) Both chaotic attractors of the uncoupled oscillators are
funnel, and the phase diffusion takes the intermediate values in
comparison with types~I and III.
In~\cite{Osipov:2003_3TypesTransitions} the values of
$D_\varphi\sim10^{-3}\div10^{-1}$ were reported for this type of
transitions. The critical curves $l_1$ and $l_2$ are supposed to be
clearly separated, both lying below the critical curve $l_3$.

(III) The uncoupled oscillators considered are characterized by
highly incoherent chaotic attractors, and the phase diffusion is
rather strong (for the system considered
in~\cite{Osipov:2003_3TypesTransitions} $D_\varphi$ exceeds
$10^{-1}$). In this case, the critical curve $l_1$ is expected to
lie above both curves $l_2$ and $l_3$, i.e., the phase
synchronization regime occurs after one of the positive Lyapunov
exponents passes to negative values.

So, Lyapunov exponents seem to be important to distinguish between
different types of transitions to/from the phase synchronization
regime. Moreover, PS regime arising is usually described in terms of
Lyapunov exponents (LEs)~(see,
e.g.~\cite{Pikovsky:2002_SynhroBook,Anishchenko:2001_SynhroBook,Boccaletti:2002_ChaosSynchro,%
Osipov:2003_3TypesTransitions}). Eventually, we have to compare our
results concerning different types of transitions from/to the PS
regime with the ones given in~\cite{Osipov:2003_3TypesTransitions}.
Therefore, let us consider the relation between critical curves
$l_1$, $l_2$ and $l_3$  corresponding to the onset of phase
synchronization, the transition of one of the zero Lyapunov
exponents to negative values and the passing of the one of the
positive Lyapunov exponents through zero for the coupled R\"ossler
oscillators~(\ref{eq:Roesslers}) on the $(\omega_d,\varepsilon)$
parameter plane, respectively (see Fig.~\ref{fgr:CriticalCurves}).

As it was mentioned above, the autonomous
systems~(\ref{eq:Roesslers}) under study with the pointed values of
control parameters (and zero coupling strength) are characterized by
the phase coherent attractors in the whole range of
$\omega$-parameter variation. Correspondingly, the phase of chaotic
signal increasing approximately uniformly, the phase diffusion
coefficient $D_\varphi$ is rather small and does not exceed the
value of $10^{-3}$ for all values of $\omega$-parameter. Therefore,
we must expect that only the first type of transitions described
in~\cite{Osipov:2003_3TypesTransitions} is observed for two
unidirectionally coupled oscillators~(\ref{eq:Roesslers}), whereas
two other types (II and III, respectively) can not be observed for
this system. Nevertheless, having considered the dynamics of two
unidirectionally coupled R\"ossler systems~(\ref{eq:Roesslers}), we
have observed types II and III of transitions as well as type I (see
Fig.~\ref{fgr:CriticalCurves}).

\begin{figure}[tb]
\centerline{\includegraphics*[scale=0.4]{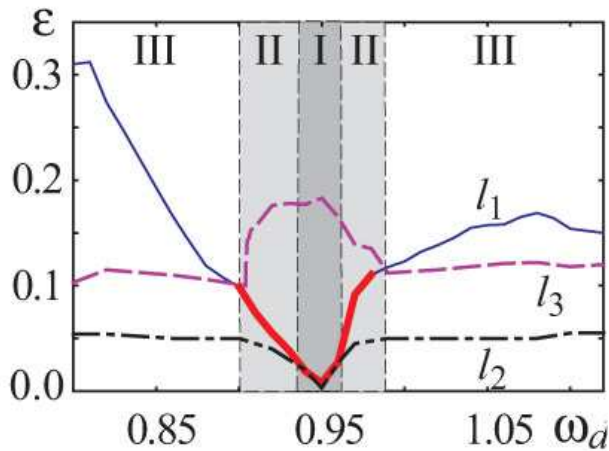}} \caption{(Color
online) The critical curves $l_1$, $l_2$ and $l_3$ on the
$(\omega_d,\varepsilon)$-plane. The critical curve $l_1$ corresponds
to the PS boundary, it is shown in the range of the small parameter
mismatch (the bold solid line) of coupled oscillators as well as in
the range of the large one (the thin solid line). The dotted line
$l_2$ corresponds to the transition of one of the zero Lyapunov
exponents to negative values. The dashed line $l_3$ taken from our
previous work~\cite{Harmov:2005_GSOnset_EPL} corresponds to zero
crossing of the one of the positive Lyapunov exponents. The
parameter plane is divided into three regions according to three
types of transitions to phase synchronization described
in~\cite{Osipov:2003_3TypesTransitions} (regions I, II and III,
respectively)} \label{fgr:CriticalCurves}
\end{figure}

As it was mentioned above, the transition of one of the zero
Lyapunov exponents to negative values is supposed to be closely
connected with the onset of PS in the case when coupled oscillators
are characterized by the
phase coherent attractors (see also~\cite{Osipov:2003_3TypesTransitions,%
Rosenblum:1997_LagSynchro}). Our results refute this statement:
Fig.~\ref{fgr:CriticalCurves} shows that the curve $l_2$ coincides
with the curve $l_1$ (onset of PS) only in a very small range of the
parameter mismatch of coupled oscillators (this range is shown in
Fig.~\ref{fgr:CriticalCurves} in dark grey color and marked by
number~I), although uncoupled both the drive and response R\"ossler
systems are characterized by the phase coherent attractors in the
whole range of the $\omega_d$-parameter variation. We can not say
that the transition of one of the zero LEs to negative values
determines PS arising even in the case of small mismatch of system
parameters. Indeed, for the parameter $\omega_d=0.9$ separating two
types of the PS destruction the coupling strength value
$\varepsilon_{PS}\approx0.099$ corresponding to the onset of PS two
times exceeds the $\varepsilon_{l_2}$-value corresponding to
$l_2$-curve ($\varepsilon_{l_2}\approx0.05$) (see also
Fig.~\ref{fgr:CriticalCurves}). Therefore, we come to conclusion
that the transition of one of the zero LEs to negative values
precedes arising of the PS regime but does not determine it. We
suppose that this transition may be connected with time scale
synchronization~\cite{Hramov:2004_Chaos,Aeh:2005_SpectralComponents},
but this problem requires further consideration.

Thus, despite the fact, that both interacting chaotic oscillators
are characterized by the phase coherent attractors for zero coupling
strength the first type of transitions described in
Ref.~\cite{Osipov:2003_3TypesTransitions} is observed only in the
narrow range of the $\omega_d$-parameter values (see
Fig.~\ref{fgr:CriticalCurves}). Moreover, two other types of
transitions, described in~\cite{Osipov:2003_3TypesTransitions} and
supposed to correspond to the synchronization of oscillators with
initially incoherent chaotic attractors, are also observed in the
considered system~(\ref{eq:Roesslers}) of oscillators with
originally coherent chaotic attractors. Indeed,
in~\cite{Osipov:2003_3TypesTransitions} it was reported that the
case when the boundary of the PS regime arising (curve $l_1$) is
between the critical curves $l_2$ and $l_3$ corresponds to the
synchronization of oscillators with funnel chaotic attractors.
Nevertheless, for the considered system~(\ref{eq:Roesslers}) this
type of transitions takes place in the $\omega_d$-parameter value
ranges shown in Fig.~\ref{fgr:CriticalCurves} in light grey color
and labeled by number~II. Similarly, the third type of transitions
described in~\cite{Osipov:2003_3TypesTransitions} and supposed to
correspond to the systems with highly incoherent attractors (the
critical curve $l_1$ lies above both the critical curves $l_2$ and
$l_3$) is also observed in~(\ref{eq:Roesslers}) despite of the
initial coherence of system attractors (white regions~III in
Fig.~\ref{fgr:CriticalCurves}).

It should also be noted that the location of the critical curve
$l_3$ on the $(\omega_d;\varepsilon)$-plane coincides with the onset
of the generalized synchronization (GS)
regime~\cite{Rulkov:1995_GeneralSynchro,%
Rulkov:1996_AuxiliarySystem,Pyragas:1996_WeakAndStrongSynchro}. The
mechanisms determining the location of the GS boundary on the
parameter plane have been considered for unidirectionally coupled
R\"ossler systems~(\ref{eq:Roesslers}) in our previous
paper~\cite{Harmov:2005_GSOnset_EPL} with the help of the modified
system approach~\cite{Aeh:2005_GS:ModifiedSystem}. Based on the
results given in~\cite{Harmov:2005_GSOnset_EPL} we can make a
decision that the location of the GS boundary (and, correspondingly,
the critical curve $l_3$) does not relate with phase synchronization
phenomenon.

Thus, we come to conclusion that for two unidirectionally coupled
R\"ossler oscillators~(\ref{eq:Roesslers}) two types of the PS
destruction are possible in dependence on the system parameter
mismatch. Of course, one can distinguish three types of the PS
destruction according to the location of the PS regime boundary and
critical curves $l_2$ and $l_3$ on the parameter plane, but there is
no reason of doing that. Contrary to distinguishing PS destruction
types between two classes (when there are well-known physical
mechanisms determining such approach), the consideration of three
types of transitions to PS based on the locus of the critical curves
$l_{1,2,3}$ on the parameter plane does not seem to be reasonable,
since the positions of $l_{2,3}$ curves are caused by different
mechanisms which do not concern the PS phenomenon.

\section{Phase synchronization destruction in the chaotic systems with initially incoherent
chaotic attractors} \label{sct:RoesslersNonCoherent}

Since the behavior of unidirectionally coupled R\"ossler systems
having been examined we came to conclusion that for the oscillators
with initially phase coherent attractors~(\ref{eq:Roesslers}) two
types (rather than three) of the PS destruction should be
distinguished, it is quite reasonable to learn what types of PS
destruction may be observed if two oscillators with initially
incoherent chaotic attractors are considered. To clarify this point
we have studied two mutually coupled R\"ossler systems
\begin{equation}
\begin{array}{l}
\dot x_{1,2}=-\omega_{1,2}y_{1,2}-z_{1,2},\\
\dot y_{1,2}=\omega_{1,2}x_{1,2}+ay_{1,2}+d(y_{2,1}-y_{1,2}),\\
\dot z_{1,2}=0.1+z_{1,2}(x_{1,2}-8.5) \label{eq:FunnelRsslrs}
\end{array}
\end{equation}
from the point of view of the concept discussed in
Sec.~\ref{sct:Roesslers} and \ref{sct:VdP}. Note, it is the system
that were considered in~\cite{Osipov:2003_3TypesTransitions}, for
which three types of transitions connected with the attractor
coherence properties and critical curves location had been described
(Fig.~\ref{fgr:FnlRsslrPSBoundary}).

\begin{figure}[tb]
\centerline{\includegraphics*[scale=0.35]{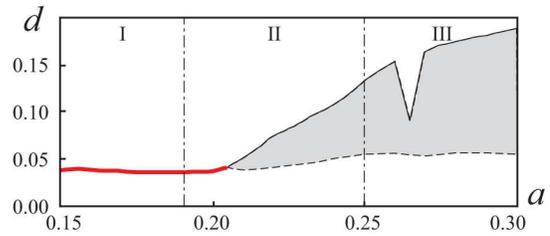}} \caption{The
phase synchronization boundary of two coupled R\"ossler
systems~(\ref{eq:FunnelRsslrs}). The bold solid line corresponds to
the phase synchronization regime destruction when both chaotic
attractors retain their phase coherence both above and below the
synchronization threshold. The thin line presents the
synchronization boundary when the PS destruction is accompanied by
break of the attractor coherence. The change of the type of the PS
regime destruction takes place at $a_*\approx0.205$. The attractor
of the first R\"ossler system is incoherent in the region shown in
grey color. The lower bound of the attractor non-coherence area is
shown by the dashed line. The areas corresponding to three types of
transitions to PS regime described
in~\cite{Osipov:2003_3TypesTransitions} are labeled by numbers I, II
and III, respectively} \label{fgr:FnlRsslrPSBoundary}
\end{figure}

The control parameter values of the systems~(\ref{eq:FunnelRsslrs})
have been selected the same as they were given
in~\cite{Osipov:2003_3TypesTransitions}. In
Eq.~(\ref{eq:FunnelRsslrs}) $d$ is the coupling strength,
$\omega_1=0.98$, $\omega_2=1.02$, the parameter $a\in[0.15;0.3]$
determines the topology of chaotic attractors. It is known, that
when $a$ exceeds the critical value $a_c$ ($a_{c1}\approx0.186$ for
$\omega_1$ and $a_{c2}\approx0.195$ for $\omega_2$) the chaotic
attractor of autonomous R\"ossler system becomes
incoherent~\cite{Osipov:2003_3TypesTransitions}. Following
\cite{Osipov:2003_3TypesTransitions} the phase $\varphi(t)$ has been
defined as the rotation angle ${\varphi=\arctan(\dot y/\dot x)}$ on
the plane ${(\dot x;\dot y)}$ to study PS of the systems with
initially funnel attractors.

Having studied the behavior of two mutually coupled R\"ossler
systems~(\ref{eq:FunnelRsslrs}) along the boundary of PS regime on
the plane $(a,d)$ given in~\cite{Osipov:2003_3TypesTransitions}, we
have also found two scenarios of the PS regime destruction described
in Sec.~\ref{sct:Roesslers} as well as for the case of two
unidirectionally coupled oscillators~(\ref{eq:Roesslers}). For small
values of $a$-parameter (${a< a_*\approx0.205}$) the first type of
the PS regime destruction takes place when chaotic attractors of
both systems keep their coherence on the plane ${(\dot x, \dot y)}$
both above and below the boundary of PS
(Fig.~\ref{fgr:FnlRsslrPSBoundary}). When $a> a_*$ the chaotic
attractor of the first system loses its coherence as soon as the
coupling strength $d$ is below the onset of PS, with the attractor
of the second system remaining to be phase coherent. Note, that
$a_*>a_{c1,2}$ therefore the change of the PS destruction type takes
place when both coupled systems are characterized by the initially
funnel chaotic attractors on the plane $(x,y)$.

To make our decision more convincing, we have also calculated the
coherence measure $\rho$ vs. coupling parameter $d$ for two
$a$-parameter values corresponding to both types of the PS regime
destruction, to be exact ${a_I=0.19<a_*}$ and ${a_{II}=0.21>a_*}$
(Fig.~\ref{fgr:FnlRsslrCoherenceMeasure}). Since the evolution of
the chaotic attractors is considered on the velocity plane $(\dot
x,\dot y)$ we have used
\begin{equation}
\label{eq:VelocityCoherenceMeasure}
\rho=\min\limits_{t\rightarrow+\infty} \sqrt{\dot x^2(t)+\dot
y^2(t)}
\end{equation}
instead of Eq.~(\ref{eq:CoherenceMeasure}).  Evidently, the chaotic
attractor of the first R\"ossler system remains phase coherent both
below and above the PS boundary for ${a_I=0.19}$ and loses its
coherence when the PS regime is destroyed in the case of
${a_{II}=0.21}$.

\begin{figure}[tb]
\centerline{\includegraphics*[scale=0.35]{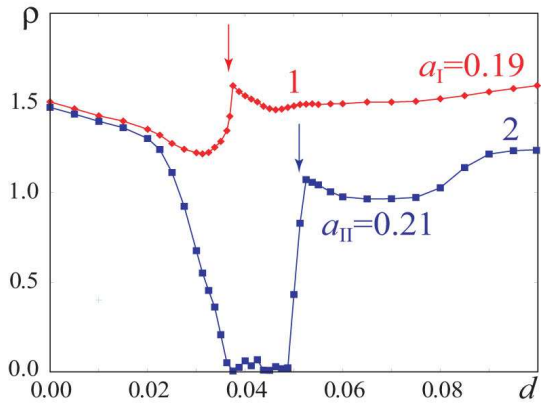}}
\caption{(Color online) The dependencies of the coherence measure
$\rho$ on the coupling strength $d$ for $a_I=0.19$ (curve 1) and
$a_{II}=0.21$ (curve 2). The arrows show the threshold of the PS
regime} \label{fgr:FnlRsslrCoherenceMeasure}
\end{figure}

The evolution of the chaotic attractor of the first R\"ossler
system, when the PS regime is destroyed, is shown in
Fig~\ref{fgr:FnlRsslrPortraits}. The first type of the PS regime
destruction, when the chaotic attractors are phase coherent both
above and below the PS boundary, is shown in
Fig~\ref{fgr:FnlRsslrPortraits},\,\textit{a,b}.
Fig.~\ref{fgr:FnlRsslrPortraits},\,\textit{c,d} illustrates the
second type of transitions accompanied by the destruction of the
coherence of one of the chaotic attractors just below the PS
boundary.

\begin{figure}[tb]
\centerline{\includegraphics*[scale=0.4]{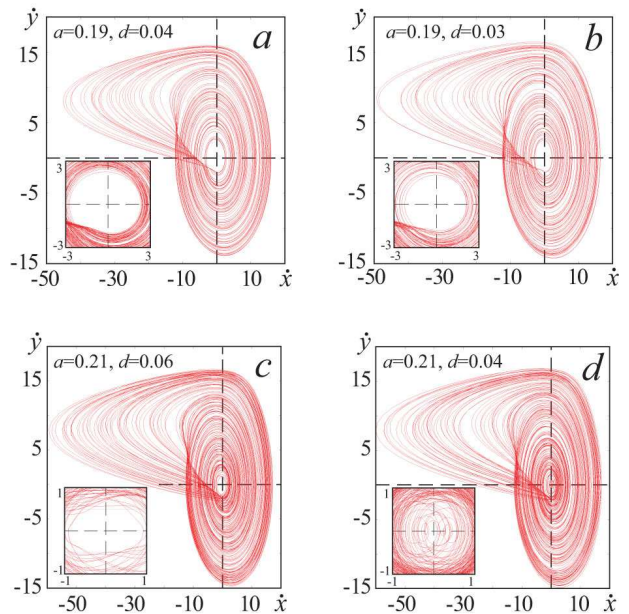}} \caption{(Color
online) Attractors of the first R\"ossler
system~(\ref{eq:FunnelRsslrs}) for two values of $a$-parameter
corresponding to two types of the PS regime destruction. Figures
(\textit{a,b}) show the evolution of the system behavior on the
velocity plane $(\dot x,\dot y)$ for $a_I=0.19$ when the attractor
remains phase coherent both above (\textit{a}) and below
(\textit{b}) the PS regime boundary. The second scenario of the PS
regime destruction ($a_{II}=0.21$) accompanied by the phase
coherence loss of the chaotic attractor is shown in
Fig.~(\textit{c,d}). The behavior of coupled oscillators is
synchronized in Fig.~(\textit{c}) when the attractor of the first
R\"ossler system on the plane ${(\dot x,\dot y)}$ is phase coherent.
Alternatively, the phase incoherent attractor is shown in
Fig.~(\textit{d}) after the PS regime has been destroyed. In the
frames the trajectories on the plane ${(\dot x,\dot y)}$ near the
origin are shown.} \label{fgr:FnlRsslrPortraits}
\end{figure}

Thus, for two coupled systems with initially funnel attractors, the
phase coherence of chaotic attractor on the velocity plane $(\dot
x,\dot y)$ is destroyed in the same way as it was observed on the
plane $(x,y)$ for the systems~(\ref{eq:Roesslers}) with initially
phase coherent attractors.

Taking into account all obtained results we come to conclusion that
for the system of two mutually coupled R\"ossler oscillators with
initially phase incoherent attractors~(\ref{eq:FunnelRsslrs}) as
well as for the oscillators with phase coherent
attractors~(\ref{eq:Roesslers}) two types (instead of three) of the
PS destruction should be distinguished. This conclusion is also
confirmed by Fig.~6 given in~\cite{Osipov:2003_3TypesTransitions}
where the mean frequency ratio ${\Omega_1/\Omega_2}$ versus coupling
strength is shown. The behavior of ${\Omega_1/\Omega_2}$ is quite
different for the first and the second types of transitions to PS
proposed in~\cite{Osipov:2003_3TypesTransitions} but is the same for
the second and the third ones.

\section{Conclusion}
\label{sct:Conclusion}

We have described two different types of the PS regime destruction
as well as the mechanisms resulting in the destruction, which are
the same as in the case of the classical synchronization of periodic
oscillators. The first of them is caused by the loss of the common
rhythm of chaotic oscillations and the second one is caused by the
loss of phase coherence of chaotic attractor. These types have been
observed in the systems with initially both phase coherent and
funnel chaotic attractors. The transition of one of the zero
Lyapunov exponents to negative values precedes arising of the PS
regime but does not cause it is the other important result.
Similarly, the passing of the one of the positive Lyapunov exponent
through zero does not relate to PS phenomenon. Therefore, there is
no reason to use these LEs for the PS onset description.

We assume that the number of the possible transitions from the phase
synchronization regime to the asynchronous dynamics might not be
limited by two types only which are described in our paper. For the
more complex systems (e.g., for the system with the larger dimension
of the phase space) the breakdown of PS may also be associated with
global bifurcations, e.g., homoclinic orbits, Takens-Bogdanov
bifurcation points,
etc.~\cite{Postnov:1999_HomoclinicBifurcationAndPS}. Nevertheless,
we suppose that the two considered types of the PS regime
destruction should be typical for a wide class of nonlinear systems,
e.g., such as Pierce diode~\cite{Filatov:2006_PierceDiode_PLA} or
laser system~\cite{Boccaletti:2002_LaserPSTransition_PRL}. In
particular, we have observed the same types of behavior in the
vicinity of the PS boundary for the case of two unidirectionally
coupled generators with tunnel diodes described in
Ref.~\cite{Rosenblum:1997_SynhroHuman}.

\section*{Acknowledgment}

We thank Dr. S.V. Eremina for English language support. This work
was supported by U.S. Civilian Research and Development Foundation
for the Independent States of the Former Soviet Union (CRDF), grant
REC--006, the Supporting program of leading Russian scientific
schools (project NSh--4167.2006.2), RFBR (07--02--00044 and
05--02--16273) and ``Dynasty'' Foundation.


\end{document}